\documentstyle[11pt,amssymb]{article}
\def\dalemb#1#2{{\vbox{\hrule height .#2pt
        \hbox{\vrule width.#2pt height#1pt \kern#1pt
                \vrule width.#2pt}
        \hrule height.#2pt}}}

\def\0{{\sst{(0)}}}
\def\1{{\sst{(1)}}}
\def\2{{\sst{(2)}}}
\def\3{{\sst{(3)}}}
\def\4{{\sst{(4)}}}
\def\5{{\sst{(5)}}}
\def\6{{\sst{(6)}}}
\def\7{{\sst{(7)}}}
\def\8{{\sst{(8)}}}

\def\ep{\epsilon}
\def\td{\tilde}
\def\wtd{\widetilde}

\let\a=\alpha \let\b=\beta

\def\nn{\nonumber} \def\bd{\begin{document}} \def\ed{\end{document}}
\def\ds{\documentstyle} \let\fr=\frac \let\bl=\bigl \let\br=\bigr
\let\Br=\Bigr \let\Bl=\Bigl 
\let\bm=\bibitem
\let\na=\nabla
\let\pa=\partial \let\ov=\overline 
\newcommand{\be}{\begin{equation}} 
\newcommand{\ee}{\end{equation}} 
\def\ba{\begin{array}}
\def\ea{\end{array}}
\def\ft#1#2{{\textstyle{{\scriptstyle #1}\over {\scriptstyle #2}}}}
\def\fft#1#2{{#1 \over #2}}
\def\del{\partial}
\def\sst#1{{\scriptscriptstyle #1}}
\def\oneone{\rlap 1\mkern4mu{\rm l}}
\def\ie{{\it i.e.\ }}
\def\via{{\it via}}
\def\semi{{\ltimes}}
\def\str{{\rm str}}
\def\jm{{\rm j}}
\def\im{{\rm i}}
\def\bOmega{{{\bar\Omega}}}
\def\Qn{{{Q_{\sst{\rm N}}}}}
\def\tX{{{\wtd X}}}

\def\mapright#1{\smash{\mathop{-\!\!\!-\!\!\!-\!\!\!-\!\!\!-\!\!\!
             \longrightarrow}\limits^{#1}}}
\def\maprightt#1#2{\smash{\mathop{-\!\!\!-\!\!\!-\!\!\!-\!\!\!-\!\!\!
             \longrightarrow}\limits^{#1}_{#2}}}

\newcommand{\ho}[1]{$\, ^{#1}$}
\newcommand{\hoch}[1]{$\, ^{#1}$}
\newcommand{\bea}{\begin{eqnarray}} 
\newcommand{\eea}{\end{eqnarray}} 
\newcommand{\ra}{\rightarrow}
\newcommand{\lra}{\longrightarrow}
\newcommand{\Lra}{\Leftrightarrow}
\newcommand{\ap}{\alpha^\prime}
\newcommand{\bp}{\tilde \beta^\prime}
\newcommand{\tr}{{\rm tr} }
\newcommand{\Tr}{{\rm Tr} }

\newcommand{\NP}{Nucl. Phys. }
\newcommand{\tamphys}{\it Center for Theoretical Physics\\
Texas A\&M University, College Station, Texas 77843}
\newcommand{\ens}{\it Laboratoire de Physique Th\'eorique de l'\'Ecole
Normale Sup\'erieure\hoch{2,3}\\
24 Rue Lhomond - 75231 Paris CEDEX 05}
\newcommand{\upenn}{\it Department of Physics and Astronomy\\
University of Pennsylvania, Philadelphia, Pennsylvania 19104}

\newcommand{\auth}{ 
H. L\"u\hoch{\dagger1} and C.N. Pope\hoch{\ddagger2}}

\begin{document}
\begin{flushright}
\hfill{CTP TAMU-26/00}\\
\hfill{UPR-899-T}\\
\hfill{hep-th/0008050}\\
\hfill{August 2000}\\
\end{flushright}

%\vspace{15pt}

\begin{center}
{ \large {\bf Branes on the Brane}}

\vspace{15pt}
\auth

\vspace{15pt}

{\hoch{\dagger}\upenn}

\vspace{15pt}
{\hoch{\ddagger}\tamphys}

\vspace{40pt}

\underline{ABSTRACT}
\end{center}

   We show that four-dimensional $N=2$ ungauged Einstein-Maxwell
supergravity can be embedded on the Randall-Sundrum 3-brane, as a
consistent Kaluza-Klein reduction of five-dimensional $N=4$ gauged
supergravity.  In particular, this allows us to describe
four-dimensional Reissner-Nordstr\"om black holes within the
Randall-Sundrum scenario.  Using earlier results on the embedding of
five-dimensional $N=4$ gauged supergravity in ten dimensions, we can
then describe the four-dimensional Einstein-Maxwell supergravity on the
3-brane, and its solutions, from a type IIB viewpoint.  We also show
that the minimal ungauged supergravities in $D=5$ and $D=6$ can be
consistently embedded in the half-maximally supersymmetric gauged
supergravities in $D=6$ and $D=7$ respectively.  These allow us to
construct solutions including BPS black holes and strings living in
``Randall-Sundrum 4-branes,'' and BPS self-dual strings living in
``Randall-Sundrum 5-branes.''  We can also lift the embeddings to
ten-dimensional massive type IIA and $D=11$ supergravity respectively.
In particular, we obtain a solution describing the self-dual string
living in the world-volume of an M5-brane, which can be viewed as an
open membrane ending on the M5-brane.

{\vfill\leftline{}\vfill
%\vskip 5pt
\footnoterule
{\footnotesize \hoch{1} Research supported in part by DOE grant 
DE-FG02-95ER40893 \vskip -12pt} \vskip 14pt
%{\footnotesize \hoch{2} Research supported in part by DOE grant
%DE-AC02-76ER03071 \vskip -12pt} \vskip 14pt
{\footnotesize  \hoch{2} Research supported in part by DOE 
grant DE-FG03-95ER40917.\vskip  -12pt}}

\pagebreak
\setcounter{page}{1}

\section{Introduction}

    The Randall-Sundrum II scenario \cite{rs2} has the intriguing feature that
despite the existence of a non-compact fifth dimension, gravity is
localised on the four-dimensional world-volume of the embedded 3-brane
wall.  This can also be understood from a Kaluza-Klein perspective,
since one can consistently replace the Minkowski metric of the 3-brane
world-volume by any Ricci-flat metric (see, for example, 
\cite{breper,hawcha,gian}), implying that pure Einstein
gravity is certainly contained within the four-dimensional theory:
%%%%%
\be
ds_5^2 = e^{-2k\, |z|}\, g_{\mu\nu}\, dx^\mu\, dx^\nu + dz^2\,.
\label{ricci}
\ee
%%%%%
(The constant $k$ will be taken to be positive throughout this paper,
since it is required for the trapping of gravity.)  It is then natural
to ask what kind of theory of gravity it is that resides on the
3-brane.  Clearly the answer to this question depends upon the nature
of the five-dimensional theory that is used in implementing the
Randall-Sundrum scenario.

   Naively, one might expect that if the five-dimensional theory is
taken to be a gauged supergravity, then the theory on the 3-brane wall
would be an ungauged supergravity with the same degree of
supersymmetry, since the AdS spacetime preserves all supersymmetry.
However, unlike an ordinary Kaluza-Klein reduction on a circle, the
fifth direction $z$ in (\ref{ricci}) is not translationally invariant,
and so in particular there will be no gauge-invariant massless
Kaluza-Klein vector arising from the reduction.\footnote{It was
recently observed in \cite{kanmyu} that if one nevertheless writes a
``standard'' reduction ansatz including a Kaluza-Klein vector, then
the inevitable lack of gauge invariance does not manifest itself until
beyond the linearised level.}  This observation is already sufficient
to show that the reduced theory in four dimensions cannot have as much
supersymmetry as the five-dimensional gauged theory.

   The situation can be clarified by looking at the Killing spinors in
an AdS background, which in horospherical
coordinates is given by $ds^2 = e^{-2kz} dx^\mu \, dx_\mu + dz^2$.
The Killing spinors are of two kinds, given by \cite{lpt}
%%%%%%
\be 
\epsilon_+ = e^{-\ft12 k\, z} \epsilon_+^0\,,\qquad \epsilon_-=
\Big(e^{\ft12 k\, z} - k\, e^{-\ft12 k\, z}\, x^\mu\, \Gamma_\mu\Big)\,
\epsilon_-^0\,,\label{killing} 
\ee 
%%%%%
where $\epsilon_\pm^0$ are
arbitrary constant spinors satisfying $\Gamma_z\, \epsilon_\pm^0 = \pm
\epsilon_\pm^0$.  In the Randall-Sundrum model, where the spacetime is
taken to be symmetric about $z=0$, one replaces $z$ by $|z|$ in
(\ref{killing}):
%%%%%%
\be 
\epsilon_+ = e^{-\ft12 k\, |z|} \epsilon_+^0\,,\qquad \epsilon_-=
\Big(e^{\ft12 k\, |z|} - k\, e^{-\ft12 k\, |z|}\, x^\mu\, \Gamma_\mu\Big)\,
\epsilon_-^0\,,\label{killing2} 
\ee 
%%%%%
Clearly half of the Killing spinors, namely $\epsilon_+$, will be
localised on the brane at $z=0$, whilst the other half, $\epsilon_-$
will not.  This implies that the four-dimensional theory on the
world-volume of the 3-brane has only half the supersymmetry of the
five-dimensional gauged supergravity.

        Thus if one considers Randall-Sundrum II in the framework of
five-dimensional simple ($N=2$) gauged supergravity, the world-volume
theory will then be four-dimensional simple ($N=1$) supergravity,
whose bosonic sector comprises only the metric.  We can also view this
as a consistent Kaluza-Klein reduction of $D=5$, $N=2$ gauged
supergravity to $D=4$, $N=1$ ungauged supergravity.  

      Four-dimensional $N=1$ supergravity does not support any BPS
$p$-branes.  The next simplest example in $D=4$ is $N=2$
Einstein-Maxwell supergravity, which admits the well-known
Reissner-Nordstr\"om black hole.  Following the previous argument, we
would expect that it can be embedded, within the Randall-Sundrum
picture, in $N=4$ gauged supergravity in $D=5$.  In section 2, we show
that it is indeed the case, and so we can construct a
Reissner-Nordstr\"om black hole in the world-volume of the 3-brane
wall.  The $N=4$ gauged supergravity itself can be obtained from a
consistent 5-sphere reduction of type IIB supergravity \cite{d5gauge},
and so this provides a consistent embedding of Einstein-Maxwell
supergravity in the world-volume of the ten-dimensional D3-brane.  The
resulting configuration can be viewed as an open string ending on the
D3-brane.

   In section 3 we generalise the procedure, to show that
six-dimensional ungauged simple supergravity can be obtained as a
consistent reduction of seven-dimensional $N=2$ gauged supergravity.
This allows us to construct a self-dual string in the world-volume of
the 5-brane wall of the seven-dimensional gauged theory.  This theory
itself can be obtained as a consistent 4-sphere reduction of $D=11$
supergravity \cite{vann,d7gauge}, and so the self-dual string can be
lifted to eleven dimensions.  The resulting configuration describes a
self-dual string living in the M5-brane, which can be viewed as an open
membrane ending on the M5-brane.  In a similar vein, we also show that
five-dimensional ungauged $N=2$ supergravity can be obtained as a
consistent reduction of six-dimensional gauged $N=2$ supergravity.  This
allows us to obtain solutions for Reissner-Nordstr\"om black holes and
strings living in the world-volume of the 4-brane wall.  Since the
six-dimensional gauged supergravity can itself be obtained as a local
$S^4$ reduction from the massive type IIA theory \cite{d6gauge}, we can
also view these solutions as living in the world-volume of the
D4/D8-brane system.

\section{Reissner-Nordstr\"om black holes in Randall-Sundrum}

\subsection{Einstein-Maxwell supergravity via Randall-Sundrum}

    Here, we show that we can obtain ungauged four-dimensional
Maxwell-Einstein ($N=2$) supergravity as a consistent Kaluza-Klein
reduction of gauged five-dimensional $N=4$ supergravity, within a
Randall-Sundrum type of framework.  The bosonic sector of the
five-dimensional theory comprises the metric, a dilatonic scalar
$\phi$, the $SU(2)$ Yang-Mills potentials $A^i_\1$, a $U(1)$ gauge
potential $B_\1$, and two 2-form potentials $A^\a_\2$ which transform
as a charged doublet under the $U(1)$.  The Lagrangian \cite{wnc}, 
expressed in the language of differential forms that we shall use here,
is given by \cite{d5gauge}
%%%%%%%%%%%%%%%
\bea
{\cal L}_5 &=& R\, {\td *\oneone} - \ft12{\td *d\phi}\wedge d\phi
-\ft12 X^4\,  {\td *G_\2}\wedge G_\2
-\ft12 X^{-2}\, ({\td *F^i_\2}\wedge F^i_\2 
+ {\td*A^\a_\2}\wedge A^\a_\2)\nn\\
& & +\fr1{2g} \epsilon_{\a\b}\, A^\a_\2\wedge dA^\b_\2 - \ft12
A^\a_\2\wedge A_\2^\a\wedge B_\1 - \ft12 F^i_\2\wedge F^i_\2\wedge B_\1\nn\\
& &  + 4 g^2\, (X^2 + 2 X^{-1})\, {\td *\oneone}\,,
\label{d5lag}
\eea
%%%%%
where $X=e^{-\ft1{\sqrt6}\, \phi}$, $F_\2^i = dA_\1^i + \ft1{\sqrt2}
g\, \ep^{ijk}\, A_\1^j\wedge A_\1^k$ and $G_\2=dB_\1$, and $\td *$
denotes the five-dimensional Hodge dual. It is useful to
adopt a complex notation for the two 2-form potentials, by defining
%%%%%
\be
A_\2 \equiv A_\2^1 + \im\, A_\2^2\,.\label{complexa}
\ee
%%%%% 

    Our Kaluza-Klein reduction ansatz involves setting the fields
$\phi$, $A_\1^i$ and $B_\1$ to zero, with the remaining metric and
2-form potentials given by
%%%%%%
\bea
ds_5^2 &=& e^{-2k\, |z|}\, ds_4^2 + dz^2\,,
\nn\\
A_\2 &=& \ft1{\sqrt2}\, e^{-k\, |z|}\, (F_\2 -\im\, {*F_\2})\,,
\label{d5ans}
\eea
%%%%%
where $ds_4^2$ is the metric and $F_\2$ is the Maxwell field of the
four-dimensional $N=2$ supergravity, and $*$ denotes the Hodge dual in
the four-dimensional metric.

    To show that this ansatz gives a consistent reduction to four
dimensions, we note from (\ref{d5lag}) that the
five-dimensional equations of motion are \cite{d5gauge}
%%%%%
\bea
d(X^{-1}\, {\td * dX}) &=& \ft13 X^4\,  {\td * G_\2}\wedge G_\2 -
\ft16 X^{-2} \, ({\td * F^i_\2}\wedge F^i_\2 + 
{\td * \bar A_\2}\wedge A_\2)\nn\\
& & - \ft{4}{3}g^2\, (X^2 - X^{-1})\, {\td * \oneone} ,\nn\\
d(X^4\, {\td * G_\2}) &=& - \ft12 F^i_\2\wedge F^i_\2 -
\ft12 {\bar A}_\2\wedge A_\2 ,\nn\\
d(X^{-2}\, {\td * F^i_\2}) &=& \sqrt{2}\, g \,
\epsilon^{ijk}\, X^{-2}\, {\td * F^j_\2}\wedge A^k_\1
- F^i_\2\wedge G_\2,\nn\\
X^{2}\, {\td * F_\3} &=& - \im \, g\, A_\2 \,,\nn\\
R_{MN} &=& 3 X^{-2}\,  \pa_M X\, \pa_N X - \ft{4}{3}g^2\,(X^2 +
2  X^{-1})\, g_{MN}\nn\\
& & + \ft12 X^4 \, (G_M{}^P G_{N P} -\ft16 g_{MN} \,
G_\2^2) + \ft12 X^{-2}\,
(F^{i\ P}_M F^{i}_{NP}  - \ft16 g_{MN}\, (F^i_\2)^2)\nn\\
& & + \ft12 X^{-2}\,  ({\bar A}_{(M}{}^P\,  A_{N)P} - \ft16
g_{MN}\, |A_\2|^2)\,,
\label{d5eom}
\eea 
where
%%%%%
\be
F_\3 =D A_\2 \equiv dA_\2  -\im\, g\, B_\1\wedge A_\2\,.\label{gauged}
\ee
%%%%% 
It follows from (\ref{d5ans}) that 
%%%%%
\be
F_\3 = -\ft1{\sqrt2}\, k\, \ep(z)\, e^{-k\, |z|}
\, (F_\2 -\im\, {*F_\2}) \wedge dz + \ft1{\sqrt2}\, e^{-k\, |z|}\,
(dF_\2 - \im\, d{*F_\2})\,,\label{f3exp}
\ee
%%%%%
where $\ep(z)=\pm1$ according to whether $z>0$ or $z<0$.
Thus the equation of motion for $F_3$ implies first of all that 
%%%%%
\be
dF_\2=0\,,\qquad d{*F_\2}=0\,,\label{maxwelleq}
\ee
%%%%%
and so then, after taking the Hodge dual of the remaining terms in
(\ref{f3exp}), we find from (\ref{d5eom}) that
%%%%%
\be
 -\ft1{\sqrt2}\, k\, \ep(z)\, e^{-k\, |z|}
\, ({*F_\2} +\im\, F_\2 ) =-\ft1{\sqrt2}\, \im\, g\,e^{-k\, |z|}\,
( F_\2 -\im\,  {*F_\2})\,,
\ee
%%%%%%
which is identically satisfied provided that
%%%%%
\be
g=  \left\{\matrix{ +k\,, & z>0\,, \cr
                    -k\,, & z<0\,. } \right. \label{kgrel}
\ee
%%%%%
Since $k$ is always positive (to ensure the trapping of gravity), this
means that the Yang-Mills gauge coupling constant $g$ has opposite
signs on the two sides of the domain wall.  This requirement is the
same as the one imposed in \cite{kalberg} for the continuity of the
Killing spinors across the boundary.  This implies, as emphasised in
\cite{dls}, that the Randall-Sundrum scenario cannot arise strictly
within the standard five-dimensional gauged supergravity, where $g$ is
a fixed parameter.  It has a completely natural explanation from a
ten-dimensional viewpoint, where $g$ arises as a constant of
integration in the solution for an antisymmetric tensor, and the
imposed $Z_2$ symmetry in fact {\it requires} that the sign must change across
the wall \cite{dls}.  For convenience, however, we shall commonly treat the
coupling constant $g$ of the gauged supergravity as if its
sign can be freely chosen to be opposite on opposite sides of the
domain wall, with the understanding that this can be justified from
the higher-dimensional viewpoint.\footnote{An alternative possibility,
developed in \cite{bergs1,kalberg}, is to introduce an auxiliary field
in the supergravity theory, with the gauge coupling constant (or mass
parameter) arising as an integration constant when the auxiliary field
is eliminated.}  

   The equations of motion for $X$ and $G_\2$ are satisfied since for
our ansatz  
%%%%%
\be
\bar A_\2\wedge A_\2 = 0 
\ee
%%%%%
and ${{\td *} A_\2}= \im\, A_\2$.  The only remaining non-trivial
equation in (\ref{d5eom}) is the Einstein equation.  In vielbein
components, the non-vanishing components of the Ricci tensor for the
metric ansatz 
%%%%%
\be
d\hat s^2 = e^{-2k\, |z|}\, ds^2 + dz^2\label{dd1met}
\ee
%%%%%
are given by
%%%%%
\bea
\hat R_{ab} &=& e^{2k\, |z|}\, R_{ab} - (D-1)\, k^2\, \eta_{ab} 
+2k\, \delta(z)\, \eta_{ab}\,,\nn\\
\hat R_{zz} &=& -(D-1)\, k^2 + 2k\, (D-1)\, \delta(z)\,,\label{ricciten}
\eea
%%%%%
where, for future reference, we have given the general expressions for a
reduction from $D$ to $(D-1)$ dimensions.  Substituting into the
five-dimensional Einstein equations, we find that the ``internal''
$(zz)$ component is identically satisfied, whilst the lower-dimensional
components imply $k^2=g^2$ (consistent with (\ref{kgrel})), and
%%%%%
\be
R_{\mu\nu} -\ft12 R\, g_{\mu\nu} = \ft12 (F_{\mu\rho}\, F_\nu{}^\rho
-\ft14 F^2\, g_{\mu\nu})\,,\label{d4einst}
\ee
%%%%%
where $R_{\mu\nu}$ is the four-dimensional Ricci tensor.
Thus we have shown that the ansatz (\ref{d5ans}), when substituted
into the equations of motion for the five-dimensional $N=2$ gauged
supergravity, gives rise to the equations of motion (\ref{maxwelleq})
and (\ref{d4einst}) of four-dimensional Einstein-Maxwell
supergravity.  

   The fact that the Kaluza-Klein reduction that we have performed
here gives a consistent reduction of the five-dimensional equations of
motion to $D=4$ is somewhat non-trivial, bearing in mind that the
five-dimensional fields in (\ref{d5ans}) are required to depend on the
coordinate $z$ of the fifth dimension.  The manner in which the
$z$-dependence matches in the five-dimensional field equations so that
consistent four-dimensional equations of motion emerge is rather
analogous to the situation in a non-trivial Kaluza-Klein sphere
reduction, although in the present case the required ``conspiracies''
are rather more easily seen.

    As in the original Randall-Sundrum model \cite{rs2}, an external
delta-function source is needed at $z=0$, to compensate the delta
function in the five-dimensional Ricci tensor given by
(\ref{ricciten}) that results from having introduced the modulus signs
in $|z|$ in (\ref{d5ans}).  Smooth gravity trapping solutions remain
elusive, and may be incompatible with supersymmetry in $D\ge 5$
\cite{kl,bc,gl}.  Of course if we omitted the modulus signs,
our ansatz (\ref{d5ans}) would satisfy the bulk supergravity equations
everywhere.\footnote{We should emphasise that from a purely mathematical
point of view all the Kaluza-Klein reductions that we consider in this
paper can be recast as fully exact consistent reductions, with no
delta-function sources needed, if we omit the modulus signs on $z$
everywhere.  There would also then be no sign-reversal of the
gauge-coupling constant $g$ on passing through $z=0$.  However, since
our goal is to describe supergravity localised on the domain wall, it
is appropriate here to introduce the modulus signs, and pay the price
of needing delta-function sources.}

    One indication of the localisation of gravity in the usual
Randall-Sundrum model is the occurrence of the exponential factor
in the metric $ds_5^2 = e^{-2k\, |z|}\, dx^\mu\, dx_\mu + dz^2$, which
falls off as one moves away from the wall.  It is therefore
satisfactory that we have found that this same exponential fall-off
occurs for the complete reduction ansatz (\ref{d5ans}), which we
derived purely on the basis of the requirement of consistency of the
embedding. 

   In our derivation of the reduction ansatz we have concentrated on
the bosonic sectors of the four-dimensional and five-dimensional
supergravities.  Since we have proved the consistency in the bosonic
sector, and since we know that the background where the
four-dimensional metric is flat admits Killing spinors (namely the
$\ep_+$ spinors given in (\ref{killing})), it follows that there must
exist a straightforward extension of our ansatz to include the
fermions.   In fact we can easily see that the exponential $z$
dependence matches properly in all the equations, with the vielbein
and gauge fields having $e^{-k\, |z|}$ factors in the reduction, while
the Killing spinors have $e^{-\ft12 k\, |z|}$ factors.

\subsection{Lifting the Einstein-Maxwell embedding to $D=10$}

    The consistent Kaluza-Klein reduction ansatz giving the embedding
of the five-dimensional $N=2$ gauged supergravity in type IIB
supergravity was derived in \cite{d5gauge}.  The internal reduction manifold 
is a 5-sphere, which can conveniently be described as a foliation by
$S^3\times S^1$ surfaces.  In this description, the unit $S^5$
is given by
%%%%%
\be
d\Omega_5^2 = d\xi^2 + \sin^2\xi\, d\tau^2 + \cos^2\xi\,
d\Omega_3^2\,,
\ee
%%%%%
where $0\le \xi\le \ft12 \pi$, $0\le \tau <2\pi$, and $d\Omega_3^2$ is
the metric on the unit 3-sphere.  The $SU(2)\times U(1)$ gauge fields
parameterise transitively-acting translations on this $S^3\times S^1$ group
submanifold, whilst the dilaton $\phi$ parametrises inhomogeneous
distortions of the 5-sphere.  Lifting our ansatz (\ref{d5ans}) to
$D=10$ is rather simple, since the scalar and $SU(2)\times
U(1)$ gauge fields vanish.  From \cite{d5gauge}, we therefore find
that (\ref{d5ans}) lifts to $D=10$ to give
%%%%%
\bea
d\hat s_{10}^2 &=& e^{-2k\, |z|}\, ds_4^2 + dz^2 + g^{-2}\, \Big(d\xi^2 
+ \sin^2\xi \, d\tau^2 + \cos^2\xi\, d\Omega_3^2\Big)\,,\nn\\
\hat H_\5 &=& 4g\, \ep_\5 + 4 g^{-5}\, \sin\xi\, \cos^3\xi\,
d\xi\wedge d\tau\wedge \Omega_\3 \,,\label{typeIIBans}\\
\hat A_\2 &=& -\ft12 g^{-1}\, \sin\xi\, e^{-k\, |z| -\im\, \tau}\,
(F_\2 -\im\, {*F_\2})\,,\nn
\eea
%%%%%
where hats denote ten-dimensional quantities, $\ep_5=e^{-4k\,
|z|}\, \ep_4\wedge dz$, with $\ep_4$ being the volume form of the
four-dimensional metric $ds_4^2$, and $\Omega_\3$ is the volume form
of the unit 3-sphere.  The complex 2-form $\hat A_\2$ is
defined by
%%%%%
\be
\hat A_\2 = A_\2^{\rm RR} + \im\, A_\2^{\rm NS}\,,
\ee
%%%%%
where $A_\2^{\rm RS}$ and $A_\2^{\rm NS}$ are the R-R and NS-NS
2-form potentials of the type IIB theory.  $\hat H_\5$ is the
self-dual 5-form of the type IIB theory, and the ten dimensional
dilaton and axion are set to zero.

\subsection{Reissner-Nordstr\"om black holes on the brane}  

   Having shown that four-dimensional ungauged Einstein-Maxwell supergravity 
arises as a consistent reduction of $N=4$ gauged five-dimensional
supergravity, which in turn is a consistent $S^5$ reduction of type
IIB supergravity, we can now take any solution of the four-dimensional
theory and lift it back to $D=5$ and $D=10$.  Examples of particular
interest are the BPS Reissner-Nordstr\"om black holes in four
dimensions, given by
%%%%%
\bea
ds_4^2 &=& -H^{-2}\, dt^2 + H^2\, dy^i\, dy^i\,,\nn\\
F_\2 &=& 2dt\wedge dH^{-1}\,,\label{bpsrn}
\eea
%%%%%
where $H(y^i)$ is any harmonic function in the transverse
3-space.\footnote{We could also, of course, consider
magnetically-charge BPS black holes, and non-extremal black holes.}
This solution lifts straightforwardly to $D=5$, using the ansatz
(\ref{d5ans}):
%%%%%
\bea
ds_5^2 &=& e^{-2k\, |z|}\, (-H^{-2}\, dt^2 + H^2\, dy^i\, dy^i) +
dz^2\,,\nn\\
A_\2  &=& \sqrt2\, e^{-k\, |z|}\, (dt\wedge dH^{-1} + \ft{\im}{2}\,
\ep_{ijk}\, \del_i H\, dy^j\wedge dy^k)\,.\label{d5bpsrn}
\eea
%%%%%
The solution could be thought of as a string, from the viewpoint of
the five-dimensional bulk theory.  Thus the solution describes an open
string ending on a D3-brane. After lifting further to type IIB
supergravity, using (\ref{typeIIBans}), we obtain
%%%%%
\bea
d\hat s_{10}^2 &=& e^{-2k\, |z|}\, (-H^{-2}\, dt^2 + H^2\, dy^i\,
dy^i) + dz^2 + g^{-2}\, \Big( d\xi^2 + \sin^2\xi\, d\tau^2 +
\cos^2\xi\, d\Omega_3^2\Big)\,,\nn\\
\hat H_\5 &=& 4g\, e^{-4k\, |z|}\, H^2\, dt\wedge d^3y\wedge dz +
4 g^{-5}\, \sin\xi\, \cos^3\xi\, d\xi\wedge d\tau\wedge \Omega_\3\,,
\label{d10res}\\
\hat A_\2 &=& -g^{-1}\, \sin\xi\, e^{-k\, |z| -\im\, \tau}\, (dt\wedge
dH^{-1} + \ft{\im}{2}\, \ep_{ijk}\, \del_i H\, dy^j\wedge dy^k)\,,\nn
\eea
%%%%%
where $k$ and $g$ are related by (\ref{kgrel}).

\section{Simple supergravities embedded in
gauged supergravities} 

   In this section we shall discuss several examples of the embedding
of an ungauged  simple supergravity in a gauged supergravity in one higher
dimension.  We already discussed the embedding of four-dimensional simple
supergravity in the introduction.  Now, we shall consider two further
examples, which are of greater interest in the sense that they contain
bosonic fields other than just the metric itself, and so they admit
BPS brane solutions.  

\subsection{Embedding minimal $D=6$ supergravity in $D=7$ and $D=11$}

     Here, we shall show how simple ungauged supergravity in $D=6$,
whose bosonic sector comprises the metric and a self-dual 3-form, can
be embedded in seven-dimensional $N=2$ gauged $SU(2)$ supergravity.
The bosonic sector of the seven-dimensional theory \cite{tv} comprises the
metric, a dilatonic scalar $\phi$, the $SU(2)$ Yang-Mills gauge
potentials $A_\1^i$, and a 3-form potential $A_\3$.   In the language
of differential forms, which we shall use here, it can be
described by the Lagrangian \cite{d7gauge}
%%%%%
\bea
{\cal L}_7 &=& R\, {{\td *}\, \oneone} -\ft12 {{\td *}\, d\phi}\wedge 
d\phi - g^2\,( \ft14 X^{-8} - 2X^{-3}
- 2 X^{2})\, {{\td *}\, \oneone}
-\ft12 X^4 \, {{\td *}\, F_\4}\wedge F_\4\nn\\
&&-\ft12 X^{-2}\, {{\td *}\, F_\2^i}\wedge F_\2^i
 +\ft12 F_\2^i\wedge F_\2^i \wedge A_\3 - \ft1{2\sqrt2} g\, F_\4\wedge
A_\3\ ,\label{d7lag}
\eea
%%%%%
where $X\equiv e^{-\phi/\sqrt{10}}$ and $F_\4=dA_\3$, together with
the self-duality condition 
%%%%%
\be
X^{4}\, {\td *F_\4} = -\ft1{\sqrt2} g \, A_\3 + \ft12
\omega_\3\,,\label{selfdual}
\ee
%%%%% 
where $\omega_\3\equiv A_\1^i\wedge F_\2^i -\ft16 g\, \ep_{ijk}
A_\1^i\wedge A_\1^j\wedge A_\1^k$.  In (\ref{d7lag}) and
(\ref{selfdual}), ${\td *}$
denotes the Hodge dual in the seven-dimensional metric.

    We find that we can consistently reduce this theory to $D=6$,
using an ansatz where $A_\1^i=0$ and $\phi=0$, together with
%%%%%
\bea
ds_7^2 &=& e^{-2k\, |z|}\, ds_6^2 + dz^2\,,\nn\\
A_\3 &=& \fft1{2k}\, e^{-2k\, |z|}\, F_3\,,\label{d7ans}
\eea
%%%%%
where $F_\3$ is a 3-form in the six-dimensional world-volume
of the 4-brane wall.  Thus $F_\4=\ep(z)\, e^{-2k\, |z|}\, F_\3\wedge
dz + 1/(2k)\, e^{-2k\, |z|}\, dF_\3$ and so substituting into
(\ref{selfdual}) we deduce that $F_\3$ must be either self-dual or
anti-self dual, and $dF_\3=0$.  Without loss of generality we shall
take $F_\3$ to be self-dual, and so we have
%%%%%
\be
g= \left\{\matrix{ +2\sqrt2 \, k\,, & z>0\,,\cr
                   -2\sqrt2\, k\,, & z<0\,. } \right.\label{d7kg}
\ee
%%%%%
We find that all the seven-dimensional equations of motion (given, in
our present notation, in \cite{d7gauge}), are satisfied provided that
$ds_6^2$ and $F_\3$ satisfy the six-dimensional equations
%%%%%
\be
R_{\mu\nu} = \ft14 F_{\mu\rho\sigma}\, F_{\nu}{}^{\rho\sigma}\,,\qquad 
F_\3 = {* F_\3}\,,\qquad dF_\3=0\,.
\ee
%%%%%
These are the bosonic field equations of six-dimensional minimal
ungauged supergravity.

    This embedding of the minimal ungauged six-dimensional supergravity 
can be lifted further, from $D=7$ to $D=11$, by making use of the
consistent $S^4$ reduction of $D=11$ supergravity.  The reduction to
the maximal $N=4$ gauged theory was obtained in \cite{vann}, and the
reduction to the half-maximal $N=2$ gauged theory that we are
considering here was constructed in \cite{d7gauge}.  In this
embedding, the internal 4-sphere is described as a foliation of $S^3$
surfaces, with the $SU(2)$ Yang-Mills potentials parameterising
transitively-acting translations on the $S^3$ group submanifold.  The
dilaton parameterises inhomogeneous deformations of the 4-sphere.
Since both $A_\1^i$ and $\phi$ vanish in our $D=7$ to $D=6$ reduction 
ansatz (\ref{d7ans}), the lifting to $D=11$ is quite simple.  Using
the results in \cite{d7gauge}, we obtain the following expressions for
the eleven-dimensional fields $d\hat s_{11}^2$ and $\hat A_\3$:
%%%%%
\bea
d\hat s_{11}^2 &=& e^{-2k\, |z|}\, ds_6^2 + dz^2 + 2 g^{-2}\, \Big(
d\xi^2 + \cos^2\xi\, d\Omega_3^2\Big)\,,\nn\\
\hat A_\3 &=& \fft1{2k}\, \sin\xi\, e^{-2k\, |z|}\, F_3 +
2\sqrt2\, g^{-3}\, \sin\xi\, (2+\cos^2\xi)\, \Omega_\3\,,
\label{d7d11}
\eea
%%%%%
where $\Omega_\3$ is the volume form of the unit 3-sphere, and again
$g$ is related to $k$ by (\ref{d7kg}).  Note that the last term just
gives a standard contribution proportional to the $S^4$ volume form
$\Omega_\4=\cos^3\xi\, d\xi\wedge \Omega_\3$ in the field strength
$\hat F_\4=d\hat A_\3$.

    Using these results, we can embed any solution of minimal ungauged
six-dimensional supergravity in a ``Randall-Sundrum 5-brane wall'' 
solution of seven-dimensional $N=2$ gauged supergravity, and then in
turn we can embed this in $D=11$ supergravity.  In particular, we can
consider a BPS self-dual string solution living in the 5-brane wall. As a
solution of the minimal six-dimensional supergravity, this is given by
%%%%%
\bea
ds_6^2 &=& H^{-1}\, (-dt^2 + dx^2) + H\, dy^i\, dy^i\,,\nn\\
F_\3 &=& dt\wedge dx\wedge dH^{-1} -\ft16 \del_i H\, \ep_{ijk\ell}\,
dy^j\wedge dy^k \wedge dy^\ell\,,
\eea
%%%%%
where $H(y^i)$ is harmonic in the transverse space.  Lifted to $D=7$
using the ansatz (\ref{d7ans}), this gives 
%%%%%
\bea
ds_7^2 &=& e^{-2k\, |z|}\, \Big(  H^{-1}\, (-dt^2 + dx^2) + H\, dy^i\,
dy^i\Big) + dz^2\,,\nn\\
A_\3 &=& \fft1{2k}\, e^{-2k\, |z|}\, \Big( 
dt\wedge dx\wedge dH^{-1} -\ft16 \del_i H\, \ep_{ijk\ell}\,
dy^j\wedge dy^k \wedge dy^\ell \Big)\,,
\eea
%%%%%
as an embedding of the self-dual string in a 5-brane wall solution of
seven-dimensional gauged supergravity.  

    This solution can then be further lifted back to $D=11$ using
(\ref{d7d11}), yielding
%%%%%
\bea 
d\hat s_{11}^2 &=& e^{-2k\, |z|}\, \Big( H^{-1}\, (-dt^2 + dx^2)
+ H\, dy^i\, dy^i\Big) + dz^2 + 2 g^{-2}\, (d\xi^2 + \cos^2\xi\,
d\Omega_3^2)\,,\nn\\
\hat A_\3 &=& \fft1{2k}\, \sin\xi\, e^{-2k\, |z|}\, \Big( 
dt\wedge dx\wedge dH^{-1} -\ft16 \del_i H\, \ep_{ijk\ell}\,
dy^j\wedge dy^k \wedge dy^\ell \Big)\nn\\
&&
 + 2\sqrt2\, g^{-3}\, \sin\xi\, (2+\cos^2\xi)\, \Omega_\3 \,,
\eea
%%%%%
where the relation between $k$ and $g$ is given in (\ref{d7kg}).
The solution can be viewed as an open membrane ending on a M5-brane.

\subsection{Embedding minimal $D=5$ supergravity in $D=6$ and $D=10$}

    In this section, we shall show that minimal ungauged
five-dimensional supergravity can be obtained as a consistent
Kaluza-Klein reduction of $N=2$ $SU(2)$-gauged supergravity in $D=6$.  

The bosonic fields in this theory comprise the metric, a dilaton $\phi$, a
2-form potential $A_\2$, and a 1-form potential $B_\1$, together with the
gauge potentials $A_\1^i$ of $SU(2)$ Yang-Mills.  The bosonic Lagrangian
\cite{romans6}, converted to the language of differential forms, is
\cite{d6gauge} 
%%%%
\bea
{\cal L}_6 &=& R\, {{\td *}\oneone} -\ft12 {{\td *}d\phi}\wedge d\phi
- g^2\Big(\ft29 X^{-6} -\ft83 X^{-2} -
2 X^2\Big)\,  {{\td *}\oneone}\nn\\
&&-\ft12 X^4\, {{\td *}F_\3\wedge F_\3} -\ft12
X^{-2}\, \Big( {{\td *}G_\2}\wedge G_\2 + {{\td *}F_\2^i}\wedge
F_\2^i \Big) \label{d6lag}\\
&& - A_\2\wedge(\ft12
dB_\1\wedge dB_\1 +\ft13 g\, A_\2\wedge dB_\1 +\ft2{27}
g^2\, A_\2\wedge A_\2 +\ft12 F_\2^i\wedge F_\2^i)\,,\nn
\eea 
%%%%%
where $X\equiv e^{-\phi/(2\sqrt2)}$, $F_\3=dA_\2$, $G_\2= dB_\1 +
\ft23g\, A_\2$, $F_\2^i = dA_\1^i + \ft12 g\, \ep_{ijk} A_\1^j\wedge
A_\1^k$, and here ${\td *}$ denotes the six-dimensional Hodge dual.

    We find that the following Kaluza-Klein ansatz gives a consistent
reduction to minimal five-dimensional ungauged supergravity.  Firstly,
we set the potentials $A_\1^i$ and $B_\1$ to zero, and also set
$\phi=0$.  The remaining fields are then taken to be
%%%%%
\bea
ds_6^2 &=& e^{-2k\, |z|}\, ds_5^2 + dz^2\,,\nn\\
G_\2 &=& \sqrt{\fft23} \, e^{-k\, |z|}\, F_\2\,,\\
A_\2 &=& \sqrt{\fft32}\, g^{-1}\, e^{-k\, |z|}\, F_\2\,,\nn
\eea
%%%%%
and the gauge-coupling $g$ is taken to be
%%%%%
\be
g = \left\{\matrix{ + \fft{3}{\sqrt2}\, k\,, & z>0\,,\cr
              -\fft{3}{\sqrt2}\, k\,,  & z <0 \,.} \right. 
\ee
%%%%%
We find that all the six-dimensional equations of motion (given, in
our notation, in \cite{d6gauge}), are then satisfied, provided that
the fields $ds_5^2$ and $F_\2$ satisfy the equations of motion of
ungauged minimal five-dimensional supergravity:
%%%%%
\bea
R_{\mu\nu} -\ft12 R\, g_{\mu\nu} &=& \ft12 (F_{\mu\rho}\, F_\nu{}^\rho -
\ft14 F^2\, g_{\mu\nu})\,,\nn\\
d{*F_\2} &=& \fft1{\sqrt3}\, F_\2\wedge F_\2\,,\qquad dF_\2=0\,.
\label{d6ans}
\eea
%%%%%

   This embedding of minimal ungauged five-dimensional supergravity on
a ``Randall-Sundrum 4-brane wall'' solution of $N=2$ gauged
six-dimensional supergravity can be further lifted to $D=10$, using
the fact that the six-dimensional gauged theory can be obtained via a
local $S^4$ reduction from the massive type IIA theory \cite{d6gauge}.
Substituting the ansatz (\ref{d6ans}) into the reduction ansatz
obtained in \cite{d6gauge}, we find
%%%%%
\bea
d\hat s_{10}^2 &=& (\sin\xi)^{\fft1{12}}\, \Big[e^{-2k\, |z|}\, ds_5^2 
 + dz^2 + 2 g^{-2}\,\Big( d\xi^2 + \cos^2\xi\,
d\Omega_3^2\Big)\Big]\,,\nn\\
\hat F_\4 &=& (\sin\xi)^{1/3}\, \Big[ \ft{20 \sqrt2}{3}\, g^{-3}\,
\cos^3\xi\, d\xi\wedge \Omega_\3 \nn\\
&&\qquad\qquad+ \ft1{\sqrt3}\, g^{-1}\, g^{-2k\,
|z|}\, {*F_\2}\wedge (\sqrt2\, \cos\xi\, d\xi - g\, \sin\xi\, dz)\Big]
\,,\\
\hat F_\3 &=& \ft1{\sqrt3}\, (\sin\xi)^{-1/3}\, g^{-1}\, e^{-k\,
|z|}\, F_\2\wedge (\sqrt2\, \cos\xi\, d\xi - g\, \sin\xi\, dz)\,,\nn\\
\hat F_\2 &=& \ft1{\sqrt3}\, (\sin\xi)^{2/3}\, e^{-k\, |z|}\, F_\2\,,\nn\\
e^{\hat\phi} &=& (\sin\xi)^{-5/6}\,,
\eea
%%%%%
where $d\hat s_{10}^2$, $\hat\phi$, $\hat F_\2$, $\hat F_\3$ and $\hat
F_\4$ are fields of the massive type IIA theory \cite{romanstype2a}, in
the notation used in \cite{d6gauge}.

   In a similar fashion to the previous examples, here we can
construct non-dilatonic black hole or string solutions in the
five-dimensional ungauged supergravity, and lift them first to
solutions in the Randall-Sundrum 4-brane wall of the six-dimensional
gauged theory, and then lift these further to $D=10$.  In ten
dimensions, the black holes or strings live in the intersection of a
D4/D8-brane system.

\section{Spacetime structure and the AdS horizon}

   A detailed analysis of the Schwarzschild black hole, embedded in
the Randall-Sundrum 3-brane wall, was carried out in \cite{hawcha}.
It was shown that it could be viewed as a black string living in the
five-dimensional AdS spacetime.  Interestingly, although gravity is
``localised on the brane'' the four-dimensional Schwarzschild black
hole has a profound influence on the spacetime geometry even out at
the AdS horizon at $z=\pm\infty$, and indeed scalar curvature
invariants diverge there \cite{hawcha}.  The solution is not expected
to be stable against ``pinching off,'' and it was argued in
\cite{hawcha} that this would happen near the AdS horizon, leading to
a stable ``black cigar.''  In this section we shall present a directly
parallel discussion for some of our examples.

   In the examples that we have considered in this paper, we can take
the $p$-brane solutions on the Randall-Sundrum wall to be
supersymmetric BPS configurations.  It is easily seen that, as in the
Schwarzschild example above, scalar curvature invariants will then
diverge on the AdS horizon.  Indeed, this can already be seen from
(\ref{ricciten}), which shows that if the Ricci tensor in the
$(D-1)$-dimensional spacetime of the brane wall is non-vanishing, then
the Ricci tensor of the $D$-dimensional bulk spacetime will diverge
exponentially as $|z|\longrightarrow\infty$.   (This is the same
degree of divergence as was encountered for curvature invariants in
\cite{hawcha}.)  For completeness, it is useful to present the full
Riemann curvature for the metric reduction ansatz (\ref{dd1met}).  In
the natural vielbein basis, the orthonormal components of the Riemann
tensor $\hat R_{ABCD}$ of the $D$-dimensional metric are related to
the components $R_{abcd}$ for the $(D-1)$-dimensional metric by
%%%%%
\bea
\hat R_{abcd} &=& e^{2k\, |z|}\,
R_{abcd} -k^2\, (\eta_{ac}\, \eta_{bd}
- \eta_{ad}\, \eta_{bc})\,,\nn\\
\hat R_{z a z b} &=& -k^2\, \eta_{ab} + 2k\, \delta(z)\,
\eta_{ab}\,.\label{hatr}
\eea
%%%%%
If we neglect the delta-function terms arising from the discontinuity
on the brane at $z=0$, we find that the scalar invariant
formed from the square of the Riemann tensor is given by
%%%%%
\be
\hat R_{ABCD}\, \hat R^{ABCD} = e^{4k\, |z|}\,
R_{abcd}\, R^{abcd} - 4k^2\, e^{2k\, |z|}\,
R + 2D(D-1)\, k^4\,,\label{hatrr}
\ee
%%%%%
where $R$ is the $(D-1)$-dimensional Ricci scalar.  Thus we see that
in general non-vanishing curvature in the $(D-1)$-dimensional metric
on the brane wall leads to exponentially-diverging curvature on the
AdS horizon, just as in \cite{hawcha}.   A non-singular solution was
constructed previously for a pp-wave propagating in AdS spacetime;
the resulting metric is the generalised Kaigorodov metric, which is
homogeneous \cite{kaig}.  A detailed analysis of pp-waves in a
Randall-Sundrum brane has been given in \cite{gibcha}.

    Following \cite{hawcha}, one can study geodesic motion in the
$p$-brane metrics living in the Randall-Sundrum walls.  We shall first
discuss the example of the BPS black holes living in the 3-brane
embedded in AdS$_5$.  If we consider a single isotropic BPS black
hole, then from (\ref{d5bpsrn}) the five-dimensional metric will be
%%%%%
\be
ds_5^2 = e^{-2k\, |z|}\, \Big[ -\Big((1+\fft{Q}{r}\Big)^{-2}\, dt^2 +
\Big(1+\fft{Q}{r}\Big)^2\, (dr^2 + r^2\, d\Omega_2^2)\Big] + dz^2\,.
\ee
%%%%%
Solving for geodesics moving in the equatorial plane, one finds that
all timelike geodesics have non-constant $z$, as do all null geodesics
that are not merely null geodesics of the four-dimensional metric.  In
the half-space $z>0$ (which we may consider without loss of
generality), we then have
%%%%%
\be
e^{-k\, z} = \left\{\matrix{ -a\,\sin k \lambda\,,& \hbox{timelike}\,, \cr
                   -a \, k\lambda\,, & \hbox{null}\,, }\right.
\ee
%%%%%
where $\lambda$ is an affine parameter and $a$ is a constant.
Associated with the $\del/\del t$ and $\del/\del\varphi$ Killing
vectors we have the two first integrals
%%%%%
\be
\fft{dt}{d\lambda} = E\, e^{2k\, z}\,
\Big(1+\fft{Q}{r}\Big)^2\,,\qquad
\fft{d\varphi}{d\lambda} = L\, e^{2k\, z}\,
\Big(1+\fft{Q}{r}\Big)^{-2}\, r^{-2}\,,
\ee
%%%%%
where $E$ and $L$ are constants.  The radial equation is
%%%%%
\be
\Big(\fft{d\td r}{d\nu}\Big)^2 + \Big[ \Big(1+\fft{\wtd
Q}{\td r}\Big)^{-2} + \Big(1+\fft{\wtd Q}{\td r}\Big)^{-4}\, \fft{\wtd
L^2}{\td r^2} -\wtd E^2 \Big] =0\,,\label{timelike}
\ee
%%%%%
where we have performed the analogous transformations of the affine
parameter to those in \cite{hawcha}, namely
%%%%%
\be
\nu = \left\{\matrix{ -\fft{1}{a^2\, k}\,\cot k\lambda\,, &
\hbox{timelike}\,,\cr
 -\fft1{a^2\, k^2\, \lambda}\,, & \hbox{null}\,,}\right.\label{affine}
\ee
%%%%%
and defined rescaled quantities
%%%%%
\be
r=a\, \td r\,,\quad Q = a\, \wtd Q\,,\quad L = a^2\, \wtd L\,,\quad E=
a\, \wtd E\,.
\ee
%%%%%
As in \cite{hawcha}, we have the somewhat intriguing result that after
performing the redefinition (\ref{affine}) of affine parameter, the
radial equation (\ref{timelike}) has reduced to the standard radial
equation for timelike geodesics in the four-dimensional metric.

      From the above results, we see that geodesics reach the horizon
at $z=\infty$ after a finite affine parameter interval, namely for
$\lambda$ approaching 0 from below.  It then follows from
(\ref{hatrr}) that the curvature diverges on the AdS horizon if
$R_{abcd}$ is non-vanishing for the metric $ds_4^2$ on the brane.
This will happen, for example, for geodesics that remain at finite
$r$ (\ie those describing bound-state orbits).  On the other hand the
geodesics that reach $r=\infty$ will have $e^{k\, z}\sim -1/(a\, k\,
\lambda)$ and $r\sim -1/(a \, k^2\, \lambda)$.  The square of the
four-dimensional Riemann tensor for the single isotropic black-hole
metric $ds_4^2$ is given by
%%%%%
\be
R^{abcd}\, R_{abcd} = \fft{8 Q^2\, (Q^2+6 r^2)}{H^8\, r^8}\,,\qquad
H=1+\fft{Q}{r}\,,
\ee
%%%%%
which goes like $48 Q^2/r^6$ as $r$ tends to infinity.  Thus we find
from (\ref{hatrr}) that as the geodesic approaches the AdS horizon,
the square of the five-dimensional Riemann tensor goes like
%%%%%
\be
\hat R^{ABCD}\, \hat R_{ABCD} \sim 40 k^4 + 48 Q^2\, a^2\, k^8\,
\lambda^2\,,
\ee
%%%%%
which therefore remains finite.  

   As in \cite{hawcha}, we can settle the question of whether these
geodesics are actually avoiding the curvature singularity on the horizon 
by looking at the components of the Riemann tensor in an orthonormal
frame parallelly propagated along a timelike geodesic.  Such a geodesic,
with $L=0$, has tangent vector $u^A$ given by
%%%%%
\be
u = -(a^2\, e^{2k\, z} -1)^{1/2}\, dz -E\, dt + H^2\, \Big(E^2 - a^2\,
H^{-2}\Big)^{1/2}\, dr
\ee
%%%%%
A unit normal vector $n^A$ that is parallelly propagated $(u^B\,
\nabla_B\, n^A=0$) along the geodesic is given by
%%%%%
\be
n=a^{-1}\,  e^{-k\, z}\, \Big(E^2 -a^2\, H^{-2}\Big)^{1/2}\, dt -
a^{-1}\, E\, e^{-k\, z}\, H^2\, dr\,.\label{normal}
\ee
%%%%%
From (\ref{hatr}) we then find that in the bulk, 
%%%%%
\be
\hat R_{ABCD}\, u^A\, n^B\, u^C\, n^D = e^{2k\, z}\, R_{abcd}\, u^a\,
n^b\, u^c\, n^d + k^2\,.
\ee
%%%%%
Using the specific form of the Riemann tensor $R_{abcd}$ for the single
isotropic black hole then gives
%%%%%
\be
\hat R_{ABCD}\, u^A\, n^B\, u^C\, n^D \sim \fft{2Q\, a\, k^2}{\lambda}
\ee
%%%%%
as $\lambda$ tends to zero, showing that the solution really does have
a curvature singularity.

   All of the above discussion closely parallels the discussion for
the Schwarzschild black hole in Randall-Sundrum in \cite{hawcha}.  In
that case, the fact that the solution is non-extremal means that it is
liable to suffer a Gregory-Laflamme \cite{greg} type of instability,
leading to the pinching-off of the five-dimensional black string to 
give a black cigar.  Since we can consider instead an extremal BPS
solution, the possibility of such an instability does not then
arise.  Thus in comparison to the Schwarzschild embedding, the
situation here is close, but no cigar.

   More generally, we can repeat the above analysis for the self-dual
string solution in the 5-brane domain wall, and the black-hole
and string solutions in the 4-brane domain wall.  We shall just
summarise the results here.  All the isotropic solutions in the various
dimensions that we have discussed in this paper take the form, in the
bulk, 
%%%%%
\be
d\hat s_D^2 = e^{-2k\, |z|} \, \Big( H^{-2\alpha}\, dx^\mu\, dx_\mu +
H^{2\beta}\, (dr^2 + r^2\, d\Omega^2_n) \Big) + dz^2\,,
\ee
%%%%%
with $H=1+Q/r^{n-1}$, where we have
%%%%%
\bea
D=5:&& \alpha=1\,,\qquad \beta=1\,,\qquad n=2\,,\qquad \hbox{black
hole} \,,\nn\\
D=6:&&\alpha=1\,,\qquad \beta=\ft12\,,\qquad n=3\,,\qquad \hbox{black
hole} \,,\nn\\
D=6:&&\alpha=\ft12\,,\qquad \beta= 1\,,\qquad n=2\,,\qquad
\hbox{string} \,,\nn\\
D=7:&&\alpha=\ft12\,,\qquad \beta= \ft12\,,\qquad n=3\,,\qquad
\hbox{self-dual string} \,.
\eea
%%%%%
For motion in the equatorial plane of the $n$-sphere, in the region
$z>0$, we shall have first integrals
%%%%%
\be
\fft{dt}{d\lambda} = E\, e^{2k\, z}\, H^{2\alpha}\,,\qquad
\fft{d\varphi}{d\lambda} = L\, e^{2k\, z}\, H^{-2\beta}\, r^{-2}\,,
\ee
%%%%%
and the $z$ equation will again give 
%%%%%
\be
e^{-k\, z} = \left\{\matrix{ -a\,\sin k \lambda\,,& \hbox{timelike}\,, \cr
                   -a \, k\lambda\,, & \hbox{null}\,, }\right.
\ee
%%%%%

    We again have that as $z$ approaches infinity, $e^{k\, z}\sim
-1/(a\, k\, \lambda)$ and $r\sim -1/(a \, k^2\, \lambda)$.  It is
easily seen that in all the cases the square of the hatted Riemann
tensor, given by (\ref{hatrr}), will again go to zero along a timelike
geodesic, as the AdS horizon is reached.  Differences emerge, however,
if we now examine the components of the hatted Riemann tensor with
respect to an orthonormal frame parallelly propagated along a timelike
geodesic.  The analogue of the normal vector $n$ in (\ref{normal}) is
%%%%%
\be
n= a^{-1}\, e^{-k\, z}\, (E^2- a^2\, H^{-2\alpha})^{1/2}\, dt -
a^{-1}\, E\, e^{-k\, z}\, H^{\alpha+\beta}\, dr\,.
\ee
%%%%%
Calculating $\hat R_{ABCD}\, u^A\, n^B\, u^C\, n^D$, we now find 
%%%%%
\be
\hat R_{ABCD}\, u^A\, n^B\, u^C\, n^D = a^2\, e^{4k\, z}\, R_{0101} \sim
\lambda^{n-3}\,,\label{res2}
\ee
%%%%%
where $R_{0101}$ denotes the vielbein component of the
$(D-1)$-dimensional Riemann tensor with 0 being time and 1 being the
$r$ direction.  Thus in those cases where the transverse space has
dimension 4, so that the $n$-sphere has $n=3$, we find that $\hat
R_{ABCD}\, u^A\, n^B\, u^C\, n^D$ remains finite rather than
diverging, as $\lambda$ tends to zero.  This occurs in the case of
black-holes living in the 4-brane in $D=6$, and self-dual strings
living in the 5-brane in $D=7$.  Although we have only exhibited one
parallelly-propagated curvature component, it seems likely that the
same feature will occur for all the components.  This is because the
$\lambda^{n-3}$ dependence seen in (\ref{res2}) simply arises from the
trade-off between the diverging $e^{4k\, z}$ factor and the $1/r^{n+1}$
fall-off of the curvature in the $(D-1)$-dimensional brane metric.

\section{Discussion and conclusion}

    In this paper, we have shown that ungauged $N=2$ supergravity
arises on the four-dimensional Randall-Sundrum 3-brane wall, obtained
as a solution of $N=4$ $SU(2)\times U(1)$ gauged supergravity in five
dimensions.  The four-dimensional supergravity emerges through a
Kaluza-Klein mechanism, as a consistent reduction from $D=5$.  Any
solution of the four-dimensional $N=2$ theory can therefore be lifted
back to five dimensions, and it acquires an interpretation as a
solution in the Randall-Sundrum wall.  In particular, the $N=2$
supergravity is large enough to admit BPS black hole solutions, and so
we can view these as living within the domain wall.  Since the
five-dimensional $N=4$ gauged supergravity can itself be
embedded in the ten-dimensional type IIB theory, via a consistent  $S^5$ 
reduction \cite{d5gauge}, the solutions can thereby be lifted to $D=10$. 

    We also considered the consistent embeddings of ungauged minimal
supergravities in gauged supergravities for a variety of dimensions.
The most immediately physically relevant example would be the
embedding of four-dimensional simple supergravity in $N=2$ gauged
five-dimensional supergravity.  Since the metric is the only bosonic
field in the four-dimensional supergravity, this example does not go
beyond previous results, such as in \cite{hawcha}, where Ricci-flat
solutions such as the Schwarzschild black hole can be viewed as living
on the 3-brane.  In this paper we considered instead the minimal
ungauged supergravities in $D=5$ and $D=6$. The former contains
gravity and a 2-form field strength in its bosonic sector, while the
latter contains gravity and a self-dual 3-form.  Both cases,
therefore, admit BPS solutions.  We showed that the $D=5$ minimal
supergravity could be embedded in six-dimensional $N=2$ $SU(2)$-gauged
supergravity, whilst the $D=6$ minimal supergravity could be embedded
in seven-dimensional $N=2$ $SU(2)$-gauged supergravity.  Thus we were
able to describe higher-dimensional examples of strings and black
holes living in ``Randall-Sundrum 4-branes,'' and self-dual strings
living in ``Randall-Sundrum 5-branes.''  In each case, a further
lifting can be performed, to $D=10$ massive IIA, and to $D=11$,
respectively.  In particular, we obtain a solution describing a
self-dual string living in the world-volume of an M5-brane.  As was
shown in \cite{warped}, half-maximal gauged supergravities can also
embedded in singular warped spacetimes in $D=11$ or $D=10$.  The BPS
back hole or string solutions we obtained in this paper can then live
in the world-volumes of intersecting M-branes or D-branes in higher
dimensions.

    We concentrated on relatively simple examples of Kaluza-Klein
domain-wall reductions in this paper, but the procedure can be
generalised to more complicated cases, with larger gauged
supergravities with more supersymmetries.

    An analysis of the spacetime structures of the embedded solutions
suggests that when the metric on the brane wall has non-vanishing
curvature, there will in general be curvature singularities at the AdS
horizons, far from the Randall-Sundrum wall.  This phenomenon was
studied in \cite{hawcha} for a Schwarzschild black hole.  Since in
that case the solution is non-supersymmetric, an instability is
expected to set in near the horizon, leading to the degenration of the
five-dimensional black string to a black cigar.  No such instability
should occur in our examples, if we take the solutions on the brane
wall to be extremal BPS $p$-branes.  The divergent curvature on the
AdS horizons can be taken as an indication that strong-coupling
effects are setting in, which would mean that the supergravity
solution would no longer be trustworthy in the region near the
horizon.  However, the analysis that we have carried out should be
valid on and near the domain wall itself, where the ungauged
supergravity is localised.

\section*{Acknowledgements}

    We are grateful to Mirjam Cveti\v c, Gary Gibbons and Kelly Stelle
for helpful discussions, and to the Theory Group at CERN for
hospitality during the course of this work.

\end{document}